\documentclass[%
prl,
 amsmath,amssymb,
preprint,%
]{revtex4-1}
\usepackage{graphicx}% Include figure files
\usepackage{dcolumn}% Align table columns on decimal point
\usepackage{bm}% bold math
\usepackage{lipsum}
\usepackage{xcolor}
\setlipsum{%
  par-before = \begingroup\color{gray},
  par-after = \endgroup
  }
\begin{document}

\preprint{AIP/123-QED}

\title[The origin of wall-shear stress fluctuations in wall-bounded turbulence]{The origin of wall-shear stress fluctuations in wall-bounded turbulence}% Force line breaks with \\
\author{Myoungkyu Lee}
\email{leemk@uh.edu}
\affiliation{Department of Mechanical Engineering, University of Houston, Houston, Texas, United States of America}

\author{Yongyun Hwang}
%\email{y.hwang@imperial.ac.uk}
\affiliation{Department of Aeronautics, Imperial College London, London, SW7 2AZ, United Kingdom}

\date{\today}% It is always \today, today,

\begin{abstract}
The origin of wall shear-stress fluctuations in wall turbulence was studied through energy dissipation at the wall. While confirming the universality in wall dissipation at small inner scales, the dissipation at larger scales is a consequence of near-wall scale interactions. In particular, the energy transport from the universal small to larger scale strengthens with Reynolds number due to the growing number of intermediate scales associated with the log layer. We anticipate that these insights broadly apply to all canonical wall-bounded turbulence for sufficiently high Reynolds numbers.
\end{abstract}

%Use showkeys class option if keyword
\maketitle
% 3750 words\\
%\section{PRL Acceptance Criteria statement}
%Our manuscript significantly advances the understanding of wall turbulence by elucidating the origins of wall shear-stress fluctuations. Our findings confirm universality in wall dissipation at small scales while demonstrating that the dissipation at larger scales stems from near-wall scale interactions, a revelation that opens new avenues within turbulence research. Importantly, we show that the energy transport from small to larger scales intensifies with the Reynolds number, highlighting scale interaction mechanisms that become more pronounced on increasing turbulence intensity. This insight, critical for the understanding of high Reynolds number turbulent flows, has broad implications for both theoretical studies and practical applications in fluid mechanics.

The precise understanding of how fluctuation energy is transported and dissipated through nonlinear interactions in fluid turbulence is one of the outstanding challenges in the physical sciences, and the related knowledge will evidently provide crucial physical insights into turbulence modeling applied for climate science \cite{Marston2016}, astrophysics \cite{Ji2006} and engineering applications \cite{Scarselli2023}. 
In wall-bounded turbulence, the large separation between viscous inner and inertial outer length scale is a distinctive feature at high Reynolds numbers (\(Re\)), and the interactions between these scales are crucially involved in the dynamical and statistical processes of the flow \cite{Hutchins2007,Marusic.2010,McKeon2017,Cho2018}. In particular, above the near-wall region, there exist a large number of hierarchically organized energy-containing motions, the size of which varies from their distance from the wall to the outer length scale \cite{Nickels.2005,Hultmark.20123vm,Hwang.2015}. On increasing \(Re\), the overall impact of these motions becomes more pronounced. It also manifests as the increased peak values in the streamwise velocity variances and the enhanced dissipation of turbulent kinetic energy \cite{GRAAFF.2000,Marusic2010,Hwang2024}. 

More than half a century ago, Townsend conjectured that the shear stress at the wall contains a slowly fluctuating part, linked with the `inactive' part of energy-containing motions in the log and outer regions \cite{Townsend.1976,Hwang2020b}. In his original work, 
the concept of inactive motions was introduced as a natural theoretical consequence of the boundary condition at the wall, which admits near-wall fluid motions in the wall-parallel directions. Over the past two decades, there has been growing evidence of the existence of such inactive motions in the near-wall region \cite{Hoyas.2006,Hwang2016c,Lee.2015wa,Deshpande2020}. 
Equipped with the accurate laboratory measurement capability and the advanced computing power, extensive research has been dedicated to understanding the influences of such inactive motions originating from the log and outer regions on the near-wall flows. Indeed, a strong correlation has consistently been observed between the large-scale components of velocity fields across the entire wall-normal locations, indicating that they may be a consequence of a direct transfer of energy from the outer to the near-wall flow \cite{Lee.20196fe}. More recent researches have leveraged the observed fluctuations in wall shear stress to propose a unified scaling for velocity variance and to extrapolate the velocity fields further from the wall \cite{Smits.2021}. Despite the important recent progress made, it has still remained elusive how these inactive motions from the log and outer regions are exactly formed in the near-wall region. As such, the precise origin of wall shear stress fluctuations in wall-bounded turbulence is yet to be understood. 

This study aims to explore the origin of wall-shear stress fluctuations associated with the inactive motions in wall-bounded turbulence. Note that the fluctuations in wall shear stress exactly represent dissipation of turbulent kinetic energy at the wall. Therefore, the origin of the wall-shear stress fluctuations must be directly associated with the processes of how turbulence is transported and subsequently dissipated in the near-wall region. To this end, we consider two distinct canonical forms of wall-bounded turbulence across various $Re$: Poiseuille flows, driven by a pressure gradient, and Couette flows, driven by boundary movement, chosen for their markedly different large-scale behaviors. The analysis utilizes the results from direct numerical simulations (DNS) of turbulent flows \cite{Lee.2015wa,Lee.20181ab}. These results are produced using the simulation software from \cite{Lee.2013kl}, which employs Fourier-Spectral methods for calculating derivatives in the streamwise ($x$) and spanwise ($z$) directions, alongside seventh-order basis spline methods for derivatives in the wall-normal ($y$) direction: for detailed information on the simulation methodology, refer to the studies by \cite{Lee.2015wa,Lee.20181ab,Lee.2013kl}. In the discussion to follow, $\langle \cdot \rangle$ denotes the average over $x$ and $z$ directions and time, and prime denotes the fluctuations. Also, the superscript ``+'' denotes normalization with friction velocity, $u_\tau$, and kinematic viscosity, $\nu$. 

It is worth highlighting the distinctions between Poiseuille and Couette flows before delving further into the analysis. 
In Couette flows, the large-scale structures in the outer region are very energetic due to the non-zero turbulence at the channel center. It was observed that a simulation domain with a streamwise length of $L_x = 100\pi\delta$ did not adequately encompass these structures present in Couette flows at $Re_\tau = 500$ where $Re_\tau$ is friction Reynolds number \cite{Lee.20181ab}. 
On the contrary, in Poiseuille flows, the production at the channel center is zero due to vanishing mean shear at the location. Therefore, the large-scale structures tend to be significantly less energetic than those in Couette flows.

\begin{figure}
  \includegraphics[width=\linewidth]{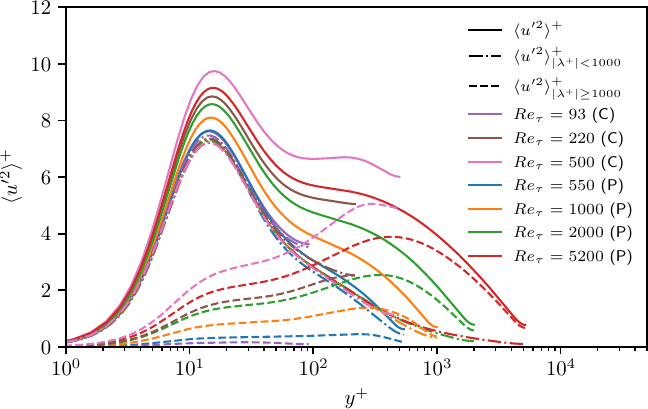}% Here is how to import EPS art
  \caption{\label{fig:uu} Streamwise velocity variances: (P) Poiseuille,  (C) Couette (------); $\langle u'^2 \rangle^+$,  ($--$) $\langle u'^2, \rangle^+_{\lambda^+<1000}$ (-- $\cdot$ --), $\langle u'^2 \rangle^+_{\lambda^+\ge1000}$.}
\end{figure}
  
\begin{figure}
  \includegraphics[width=\linewidth]{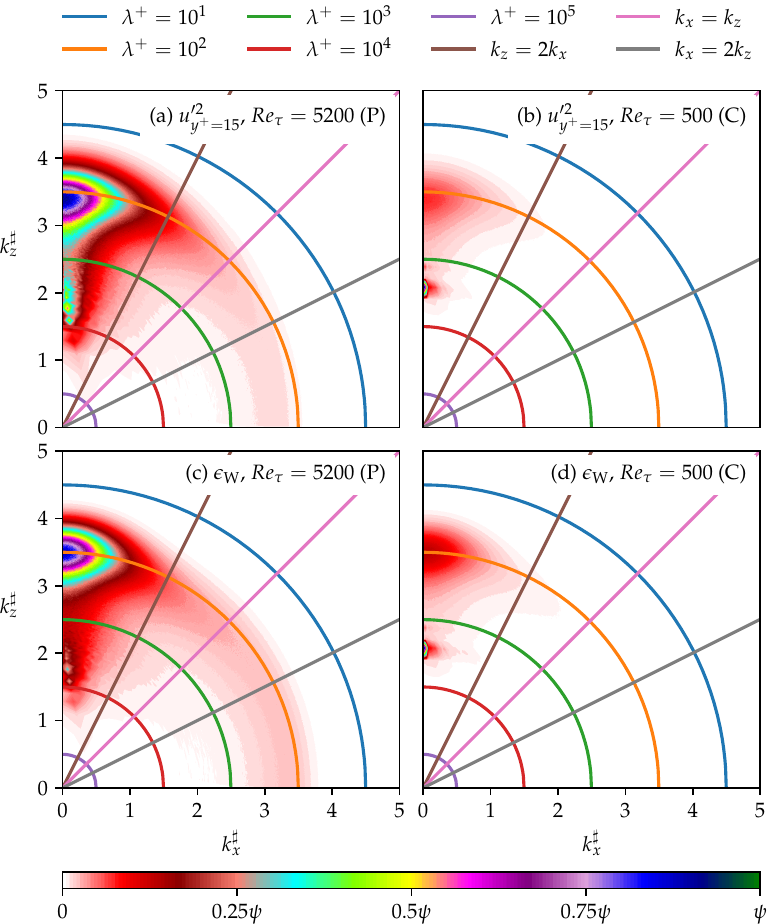}
  \caption{\label{fig:uu15_E_w} Two-dimensional spectral densities in polar-log coordinates of $u'^2$ and $\epsilon_W$: 
  (a) $E_{u'^2}^\#$ at $y^+ = 15$, $Re_\tau = 5200$, Poiseuille, $\psi = 12.1$; 
  (b) $E_{u'^2}^\#$ at $y^+ = 15$, $Re_\tau = 500$, Couette, $\psi = 144.5$; 
  (c) $E_{\epsilon_W}^\#$, $Re_\tau = 5200$, Poiseuille, $\psi = 0.68$;
  (d) $E_{\epsilon_W}^\#$, $Re_\tau = 500$, Couette, $\psi = 4.33$.
  }
\end{figure}

We start by showing the wall-normal profiles of streamwise turbulent velocity fluctuations, denoted as $\langle u'^2\rangle$, in figure~\ref{fig:uu} with solid lines. These profiles consistently display peaks around $y^+ \approx 15$, with the magnitude escalating with $Re_\tau$. The two-dimensional spectral density of $\langle u'^2 \rangle$ at $y^+ = 15$ are presented in figures~\ref{fig:uu15_E_w}(a,b) employing a polar-logarithmic format to scrutinize the isotropy and the contributions from the $k_x=0$ or $k_z=0$ modes ($k_x$ and $k_z$ represent the wavenumbers in the $x$ and $z$ directions, respectively) \cite{Lee.20196fe}.
For a two-dimensional spectral density of a statistical quantity as a function of $k_x$ and $k_z$, $E(k_x,k_z)$, the rescaled spectral density in the polar-logarithmic coordinates, $ E^\#$, is given by
\begin{subequations}
\begin{equation}
  E^\#(k_x^\#, k_z^\#) = k^2 E(k_x,k_z) /\xi,
\end{equation}
where
\begin{equation}
  k_x^\# = \xi k_x / k, \quad  k_z^\# = \xi k_z / k, \quad  k = \sqrt{k_x^2 + k_z^2}
\end{equation}  
\end{subequations}
Here, $k_\mathrm{ref}$ is an arbitrary reference wavenumber with $k_\mathrm{ref}^+=1/50 000$, and $\xi = \ln (k/k_\mathrm{ref})$. Note that the integration of $E^\#$ over $k_x^\#$ and $k_z^\#$ remains same as the integration of $E$ over  $k_x$ and $k_z$.

The spectral densities of turbulent kinetic energy, denoted by $E^\#_{u'^2}$ for both Poiseuille and Couette flows, exhibit a good scale separation at a normalized wavelength, $\lambda^+ = 1000$ ($\lambda = 2\pi/k$), as shown in figures~\ref{fig:uu15_E_w}(a,b). For convenience, we shall refer to the velocity field for $\lambda^+< 1000$ and $\lambda^+>1000$ as large and small scale, respectively. By applying a high-pass filter with the cut-off threshold $\lambda^+ = 1000$ to the $\langle u'^2 \rangle$ data, a universal behavior in the small scale is observed in both Poiseuille and Couette flows, persisting up to $y^+ \approx 70$ for various $Re$s. 
We note that the identified filtering threshold would be applicable to other canonical wall-bounded turbulent flows, such as pipe \cite{Ahn.2015,Pirozzoli.2021vy8,Yao.2023} and boundary layers flows \cite{Sillero.2013,Samie.2018}, with an expectation to produce the same turbulence statistics at the small scale.

\begin{figure}
  \includegraphics[width=\linewidth]{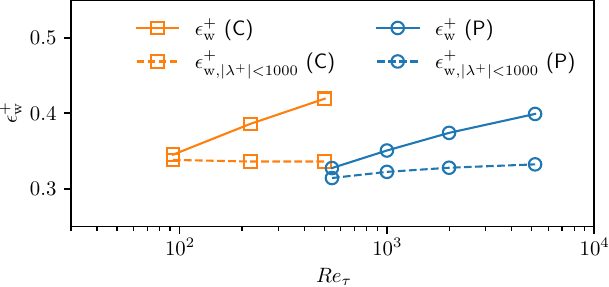}
  \caption{\label{fig:Euu_wall} Dissipation of $\langle u'^2 \rangle^+$, $\epsilon_W$ at the wall: (------) Total dissipation rate, (-- -- --) High-pass filtered dissipation rate, (P) Poiseuille,  (C) Couette.}
\end{figure}

Figures~\ref{fig:uu15_E_w}(c,d) show the spectral density of the dissipation rate of $\langle u'^2 \rangle$ at the wall, denoted by $E^{\epsilon_W}_{u'^2}$:
\begin{equation}
  E^{\epsilon_W}_{u'^2} = 2 \nu \left\langle\frac{\partial \hat{u}'}{\partial y}\frac{\partial \hat{u}'^*}{\partial y}\right\rangle_{y=0},
\end{equation}
where $\hat{u}'$ represents the Fourier transform of $u'$, and $\hat{u}'^*$ is its complex conjugate. The spectral density of wall dissipation closely resembles that of $\langle u'^2 \rangle$ at $y^+=15$, since $u'\sim \partial u'/\partial y|_{y=0}$ in the vicinity of the wall, and is confirmed in figures~\ref{fig:uu15_E_w}.  The significance of this resemblance is thoroughly discussed in \cite{Smits.2021}. Analogous to the trend observed for $\langle u'^2 \rangle$, the dissipation at small scale is therefore expected to exhibit asymptotically $Re$-invariant characteristics, with the filtering threshold of $\lambda^+ = 1000$, as shown in figure~\ref{fig:Euu_wall}. This behavior suggests that the dissipation at the near-wall small scale is at least statistically unaffected by the large scale defined here. It also indicates universality in the dissipation rate at the small scale across different flow types, such as Poiseuille and Couette flows.

Now, our focus will shift to the dissipation rate at large scale ($\lambda^+ > 1000$), highlighting the differences between Poiseuille and Couette flows at various $Re$s. There has been evidence that the large-scale turbulence at $y^+ = 15$ is directly transferred from outer flows \cite{Lee.20196fe}. As such, one could assume that the wall dissipation rate at large scale is the effect of footprint of large-scale structures in the log and outer regions. Indeed, the large-scale streamwise dissipation at the wall in figure~\ref{fig:uu15_E_w} shows a strong resemblance to the corresponding velocity fluctuations at $y^+ = 15$.

To examine these observations from the viewpoint of turbulent energetics, here we further study the transport of $\langle u'^2 \rangle$ in the near-wall region. It is worth mentioning that the terms related to pressure fluctuation have been found negligible for large scale in the near-wall region up to $y^+ = 100$. Furthermore, the production, which involves the mean velocity gradient, has a minimal effect below $y^+ = 10$ at large scale \cite{Lee.20196fe}. Consequently, the dissipation rate at  large scale is found to be balanced solely by viscous and turbulent transport terms.

First, we obtain the filtered statistics of viscous transport term, $D_{u'^2}$ and turbulent transport term, $T_{u'^2}$ from their two-dimensional spectral densities, defined below: 
\begin{subequations}
  \begin{eqnarray}
    E_{u'^2}^D &=& \displaystyle \nu \left(-k^2 \hat{u}' \hat{u}'^* + \frac{\partial^2 \hat{u}' \hat{u}'^*}{\partial y^2} \right),\\
    E_{u'^2}^T &=& \displaystyle -\left(\hat{u}'^* \widehat{\frac{\partial u_k' u'}{\partial x_k}} + \hat{u}'\widehat{\frac{\partial u_k' u'}{\partial x_k}}^* \right).
  \end{eqnarray}
\end{subequations}
\begin{figure}
  \includegraphics[width=\linewidth]{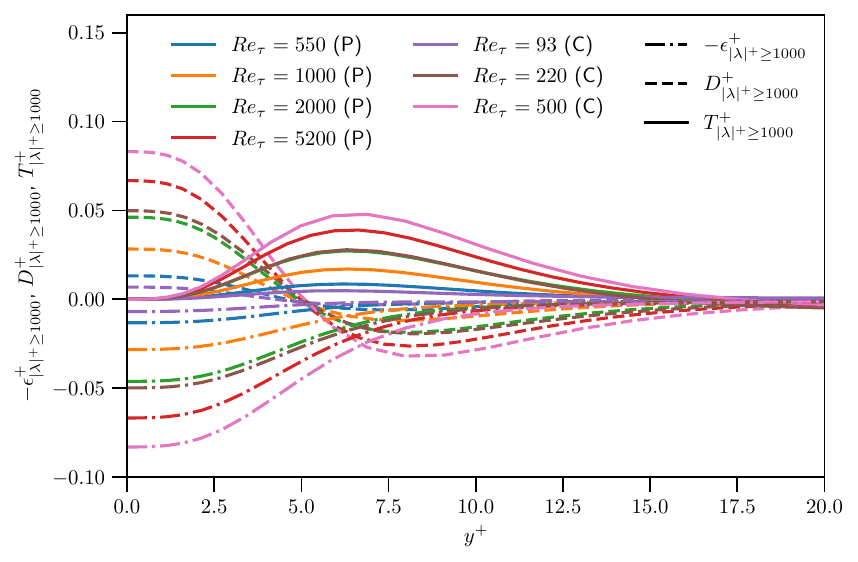}% Here is how to import EPS art
  \caption{\label{fig:Duu_Tuu} Viscous transport, $D^+$, and turbulent transport, $T^+$, of $\langle u'^2 \rangle^+$ at large scale ($\lambda^+\geq 1000$): (P) Poiseuille, (C) Couette (------); $T^+$ (- - -), $D^+$ (-- $\cdot$ --) $-\epsilon^+$.
  %WOULD BE NICER IF EPSILON IS PLOTTED.
  }
  \end{figure}
Here, $\int_0^{2h} E_{u'^2}^D \mathrm{d}y = 0$ for all $(k_x,k_z)$. Figure~\ref{fig:Duu_Tuu} shows that $D_{u'^2}$ and $T_{u'^2}$ at large scale increase with \(Re\). This highlights that in Couette flow, the motions at large scale would exert a significantly stronger influence on \(D_{u'^2}\) and \(T_{u'^2}\) despite the \(Re_\tau\) values considered substantially lower than those in Poiseuille flow. Importantly, at the wall, \(D_{u'^2} - \epsilon_{u'^2} = 0\), as all the other terms in the budget equation are null. Consequently, the energy dissipated at the wall must be transported by the viscous term from the vicinity of the wall, with its majority originating from turbulent transport. Thus, the energy at large scale, transported by turbulence, constitutes the primary source of wall dissipation especially at high $Re$s.

Energy at large scale in the near-wall region can be acquired via two different pathways. One involves the transfer of energy from small to large scale \cite{Cho2018,Kawata.2018}, while the other is the energy transport within the large scale from outer flow \cite{Lee.20196fe}. To quantitatively assess the contribution of each mechanism, we further decompose the two-dimensional spectral density of turbulent transport, $E_{u'^2}^T$, into
\begin{subequations}\label{eq:4}
  \begin{eqnarray}
    E_{u'^2}^{T^\bot} &=& -\frac12 \left(  \frac{\partial \hat{u}'^* \widehat{u'v'}}{\partial y} + \frac{\partial \hat{u}' \widehat{u'v'}^*}{\partial y} \right), \\
    E_{u'^2}^{T^\|} &=& E_{u'^2}^{T} - E_{u'^2}^{T^\bot}.
  \end{eqnarray}    
\end{subequations}
Note that $\int_0^{2h} E_{u'^2}^{T^\bot} \mathrm{d}y = 0$ for all $(k_x,k_z)$, indicating that there is no net turbulent transport in the wall-normal direction across the wavenumbers (or the two scales). Similarly, $\iint_0^{\infty} E_{u'^2}^{T^\|} \mathrm{d}k_x \mathrm{d}k_z = 0$ for all $y$, showing that there is no net inter-scale transfer by turbulence at any given wall-normal distance. Hence, $E_{u'^2}^{T^\bot}$ represents the turbulent transport in the wall-normal direction within each scale, while $E_{u'^2}^{T^\|}$ signifies the inter-scale energy transfer at given wall-normal distances. 

\begin{figure}
  \includegraphics[width=\linewidth]{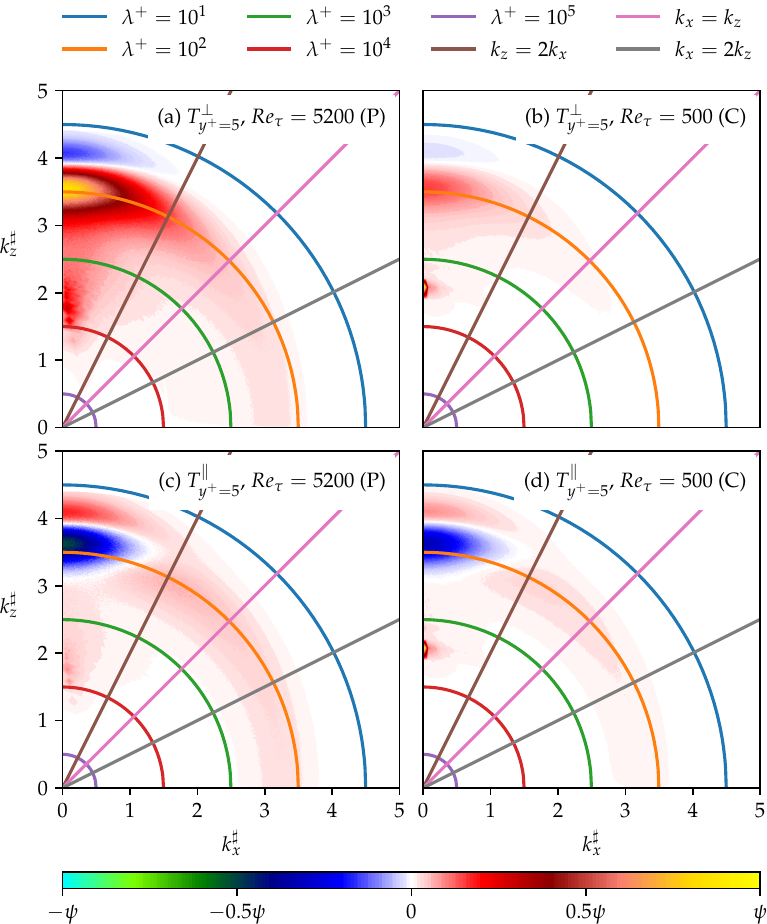}
  \caption{\label{fig:FIG4} Two-dimensional spectral densities in polar-log coordinates of $T^+_\bot$ and $T^+_\|$ at $y^+ = 5$:
  (a) $E_{T^\bot}^\#$, $Re_\tau = 5200$, Poiseuille, $\psi = 0.34$;
  (b) $E_{T^\bot}^\#$, $Re_\tau = 500$, Couette, $\psi = 1.38$;
  (c) $E_{T^\|}^\#$, $Re_\tau = 5200$, Poiseuille, $\psi = 0.34$;
  (d) $E_{T^\|}^\#$, $Re_\tau = 500$, Couette, $\psi = 1.38$}.
\end{figure}

The spectral densities of $E_{u'^2}^{T^\bot}$ and $E_{u'^2}^{T^\|}$ at $y^+ = 5$ are shown in Figure~\ref{fig:FIG4}. Both Poiseuille and Couette flows exhibit a positive region around $k_x^\# \approx 0$ and $k_z^\# \approx 2$. This suggests the energy at these wavenumbers is a consequence of turbulent transport within the large scale defined here and of that from the small scale at the given $y^+$. The amount of energy transferred from small scale is also comparable to that transported by wall-normal transport within large scale in both Poiseuille and Couette flows.

\begin{figure}
  \includegraphics[width=\linewidth]{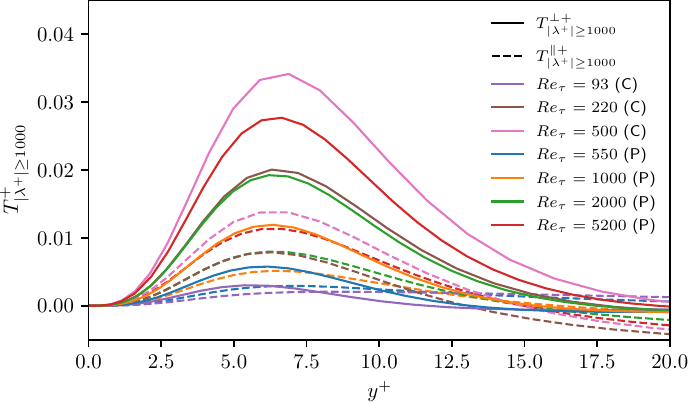}% Here is how to import EPS art\
  \caption{\label{fig:FIG5} Turbulent wall-normal transport, $T^+_\bot$, and turbulent scale transfer, $T^+_\|$ of $\langle u'^2 \rangle^+$ by large scale ($\lambda^+\geq 1000$): (P) Poiseuille,  (C) Couette; (------) $T^+_\bot$,  (- - -) $T^+_\|$.}
\end{figure}

Figure~\ref{fig:FIG5} shows the integrated values of the spectral densities of turbulent transport at large scale, $T_\bot$ and $T_\|$. In both Poiseuille and Couette flows, the values of $T_\bot$ and $T_\|$ exhibit an increase as the $Re_\tau$ rises. Notably, the data reveal that $T_\| \approx 0.4T_\bot$ at their peak values across all cases. This observation suggests that approximately 30\% of the energy at large scale in the near-wall region originates from small scale at the same wall-normal location, while the remaining 70\% is from the log and outer regions. Furthermore, this pattern is consistent in both Poiseuille and Couette flows, spanning the entire range of $Re$s investigated.

Given that the energy at the large scale defined here increases with $Re$, the increase of $T_\bot$ may not be unexpected. The increase of $T_\|$, however, deserves a further discussion, since the turbulence statistics and spectra of $u'$ at small scale remain invariant with $Re$. In the Fourier space, the spectral density of turbulent transport involves a convolution integral across the entire wavenumbers: i.e.
\begin{eqnarray}\label{eq:5}
&&E_{u'^2}^{T}(k_x,k_z) \sim \\
&&\hat{u}^*(k_x,k_z)\iint_0^{\infty}\hat{u}(k_x-s_1,k_z-s_2)\frac{\partial \hat{u}_k(s_1,s_2)}{\partial x_k} \textrm{d}s_1\textrm{d}s_2.  \nonumber
\end{eqnarray}
This is also true for both $E_{u'^2}^{T^\bot}$ and $E_{u'^2}^{T^\|}$, given its definition in Eq. (\ref{eq:4}). Therefore, even if $\hat{u}(k_x,k_z)$ for $\lambda^+<1000$ does not change with $Re_\tau$ (figures \ref{fig:uu} and \ref{fig:uu15_E_w}), $E_{u'^2}^{T^\bot}$ and $E_{u'^2}^{T^\|}$ still depend on $\hat{u}(k_x,k_z)$ for $\lambda^+>1000$, the integrated energy of which over the given range of the wavenumbers increases with $Re$. Furthermore, given the recent evidence that $\hat{u}_k(k_x,k_z)$ at $k_x \delta \sim O(1)$ and $k_z\delta \sim O(1)$ becomes smaller with $Re$ \cite{Hwang2024},  Eq. (\ref{eq:5}) implies that the increase of both $T_\bot$ and $T_\|$ at large scale is due to the increase in the range of wavenumbers that $\hat{u}(k_x,k_z)$ for $\lambda^+<1000$ can interact with. In this respect, it is finally worth mentioning that the observation of $T_\| \approx 0.4T_\bot$ in both Poiseulle and Couette flows is presumably a consequence of the definition of $T_\|$ and $T_\bot$ in (\ref{eq:4}) and the given threshold wavenumber $\lambda^+ \simeq 1000$. Therefore, this feature is also expected to be universal across different flows and $Re$s. 

In summary, the current study confirms the universal quantitative characteristics in small-scale motions of streamwise velocity variance and its dissipation at the wall, observed across both Couette and Poiseuille flows. Importantly, the wall dissipation of energy at large scale can be traced back to the scale interactions there. In particular, we found that the energy transport from the universal small scale strengthens, as $Re$ is increased, and this is a consequence of the increasing separation between the inner and outer scales, resulting in a growing number of intermediate scales associated with the log layer. 
We anticipate that these insights are broadly applicable to all canonical wall-bounded turbulence, such as zero-pressure gradient boundary layers and pipe flows, across a wide range of $Re$. However, the underlying `dynamical' mechanisms of the observed phenomena currently remain elusive. Consequently, further investigations into the dynamics of large-scale structures in near-wall flows at high $Re$ are required. 

This research used resources of the Argonne Leadership Computing Facility, a U.S. Department of Energy (DOE) Office of Science user facility at Argonne National Laboratory and is based on research supported by the U.S. DOE Office of Science-Advanced Scientific Computing Research Program, under Contract No. DE-AC02-06CH11357. Y. H. gratefully acknowledges financial support from the European Office of Aerospace Research and Development (FA8655-23-1-7023; Program Manager: Dr D. Smith).

\nocite{*}
\end{document}